\documentclass[5p]{elsarticle}
\usepackage{amsmath}
\usepackage{graphicx}
\usepackage{dcolumn}

\begin{document}
\begin{frontmatter}

\title{A simple approximation for the distribution of ions between charged plates in the weak coupling regime}
\author{A. Bazarenko}
\address{ 
University of Vienna, Computational Physics Group, Sensengasse 8/9, 1090 Vienna, Austria
}%
\author{M. Sega}%
 \ead{m.sega@fz-juelich.de}
\address{ 
University of Vienna, Computational Physics Group, Sensengasse 8/9, 1090 Vienna, Austria
}%
\address{ 
Forschungszentrum J\"ulich,
Helmholtz-Institut Erlangen-N\"urnberg,
F\"urther Stra{\ss}e  248,
90429 N\"urnberg,
Germany
}%
\begin{abstract}
The solution of the Poisson--Boltzmann equation for counterions confined between two charged plates is known analytically up to a constant, namely, the ion density in the middle of the channel.
This quantity is relevant also because it gives access, through the contact theorem, to the osmotic pressure of the system.
Here we compare the values of the ion density 
obtained by numerical and simulation approaches, 
and report a useful analytic approximation for the 
weak coupling regime in the absence of added salt, 
which predicts the value of the ion density in the 
worst case within 5\%. The inclusion of higher order terms in a Laurent 
expansion can further improve the accuracy, at the expense of simplicity.
%
\end{abstract}


\end{frontmatter}

\section{Introduction}
The release of weakly bound counterions from solid surfaces, colloids, or macromolecules, when these are put in contact with a solvent like water, is a common phenomenon that plays a key role in determining the properties of both the solvated objects and of the solvent itself. 
The problem of determining the counterion distribution between two infinite charged plates has been tackled in its many aspects (weakly and strongly correlated regimes, multivalent ions\cite{moreira01a}, presence of specific interactions\cite{ben-yakov09}, dielectric boundary conditions\cite{Kanduc07}, varying solvent per\-mit\-ti\-vi\-ty\cite{abrashkin07a}) using different approaches, ranging from mean field approaches to field-theoretical methods\cite{netz01a} and, of course, using computer simulations techniques\cite{moreira02a,Smiatek2009}.
\let\thefootnote\relax\footnote{\copyright 2018. This manuscript version is made available under the CC-BY-NC-ND 4.0 license http://creativecommons.org/licenses/by-nc-nd/4.0/}

At the mean-field level, also known as Poisson--Boltz\-mann regime, the statistical mechanic problem of determining the distribution of ions is simplified by assuming that each ion is influenced only by the average charge distribution, therefore disregarding further correlation effects. In this limit, the Poisson equation of electrostatics $\nabla^2\phi=-e \rho / \epsilon_r$, which relates the electrostatic potential $\phi$ to the charge density $e\rho$ (here we assume monovalent counterions of charge $e$) and the dielectric constant $\epsilon_r$, is combined with the Boltzmann distribution of the counterions $\rho=\rho_0 \exp(-e\phi/k_BT)$, where $k_B$ is Boltzmann's constant, $T$ is the absolute temperature, and $\rho_0$ is a normalization constant, corresponding by construction to the number density where $\phi=0$.

In the problem of counterions (no added salt) confined between two charged plates placed at distance $L$ from each other,  the potential can be chosen to be zero in the middle of the channel, so $\rho_0$ represents the number density of counterions in the middle of the channel. The solution of the Poisson--Boltzmann equation for this problem is\cite{engstrom78,Andelman95}
\begin{equation}
\rho(z) = \frac{\rho_0}{cos^2(kz)},\label{PBsol}
\end{equation}
where $k^2 = e^2 \rho_0 / (2 \epsilon_r k_BT) $.
Here $z$ is the coordinate orthogonal to the plates. 

In order to determine $\rho_0$, at least implicitly, one has to impose the normalization condition $2\sigma=\int_{-L/2}^{L/2} \rho(z) dz$, where $\sigma$ is the surface density of co-ions on each of the two plates (the surface charge of each plate being $e\sigma$). As a result, $\rho_0$ can be expressed in terms of all other parameters $e$, $\epsilon_r$, $L$ and $k_BT$, through the nonlinear equation 
\begin{equation}
k \sigma  = \rho_0 \tan(kL/2).\label{nonlin1}
\end{equation}
This equation ($k$ being a function of $\rho_0$) has no general analytic solution.
In order to make quantitative predictions for the counterion distribution, one cannot prescind from the knowledge of $\rho_0$. Therefore, one can in principle resort  to determine it numerically. Here, we show that the numerical solution is, in fact, in excellent agreement with molecular dynamics simulations results over a wide range of parameters, and we provide an analytic approximation for $\rho_0$ in the weak coupling regime, which proves to be surprisingly effective given its simplicity.
\section{Results and Discussion}
There are three length scales that are relevant for the discussion of the problem, namely (a) the interplate distance $L$, (b) the Bjerrum length\cite{bjerrum1909a} $\lambda=e^2 /(4\pi\epsilon_r k_BT)$, which is  
the distance between two ions where their electrostatic energy is the same as the thermal one, $k_BT$, and (c) the Gouy-Chapman length\cite{chapman1913li} $\mu=1/(2\pi\lambda\sigma)$, which is the distance of an ion from a charged plate at which the electrostatic energy is, again, equal to $k_BT$. It is convenient to express all distances in units of the interplate distance $L$ so that, for example, $\lambda =  \lambda^* L $, $\mu = \mu^* L$, $k = k^* / L $. In these units, Eq.~(\ref{nonlin1}) can be rewritten as 
\begin{equation}
1 = k^* \mu^* \tan(k^*/2),\label{nonlin2}
\end{equation}
where we have made use of the identity $\rho_0^* = k^{*2}\mu^*\sigma^*$. Once the (numerical) solution $k^*(\mu^*)$ is known, it is trivial to obtain the density $\rho^*(\mu^*)$. It should be noted that, by using reduced units,  $k^*$ depends only on $\mu^*$, while $\rho^*$ depends on $\mu^*$ and $\lambda^*$, or, equivalently, $\lambda^*$  and $\sigma^*$. As we already mentioned, Eq.~(\ref{nonlin2}) has no general analytical solution, but its asymptotic behavior for large and small values of $\mu^*$ is known\cite{netz01a}: 
\begin{equation}
\begin{array}{lr}
k^{*2} \approx 2/\mu^* -1/(3\mu^{*2}) + \mathcal{O}(1/\mu^{*3})& \mathrm{if~}\mu^*\gg 1 \label{limit}\\
k^{*2} \approx \pi^2 + \mathcal{O}(\mu^*) & \mathrm{if~}\mu^* \ll 1.
\end{array}
\end{equation}
These asymptotic behavior will prove to be useful to identify an approximate expression for the solution $k^*(\mu^*)$.

The validity of the Poisson--Boltzmann approximation is not always guaranteed, and depends mainly on the  coupling parameter $\Xi=2\pi \lambda^{*2} \sigma^*$.~\cite{Boroudjerdi2005}
The condition under which the Poisson--Boltzmann approximation yields correct solutions is $\Xi < \min\{1,-(\mu^*\log \mu^*)^{-1}\}$\cite{netz01a}. 
In this work we restrict our Molecular Dynamics (MD) simulations  to parameter sets that always satisfy this inequality.

\begin{figure}[t!]
\includegraphics[width=\columnwidth,trim={35 5 30 10},clip]{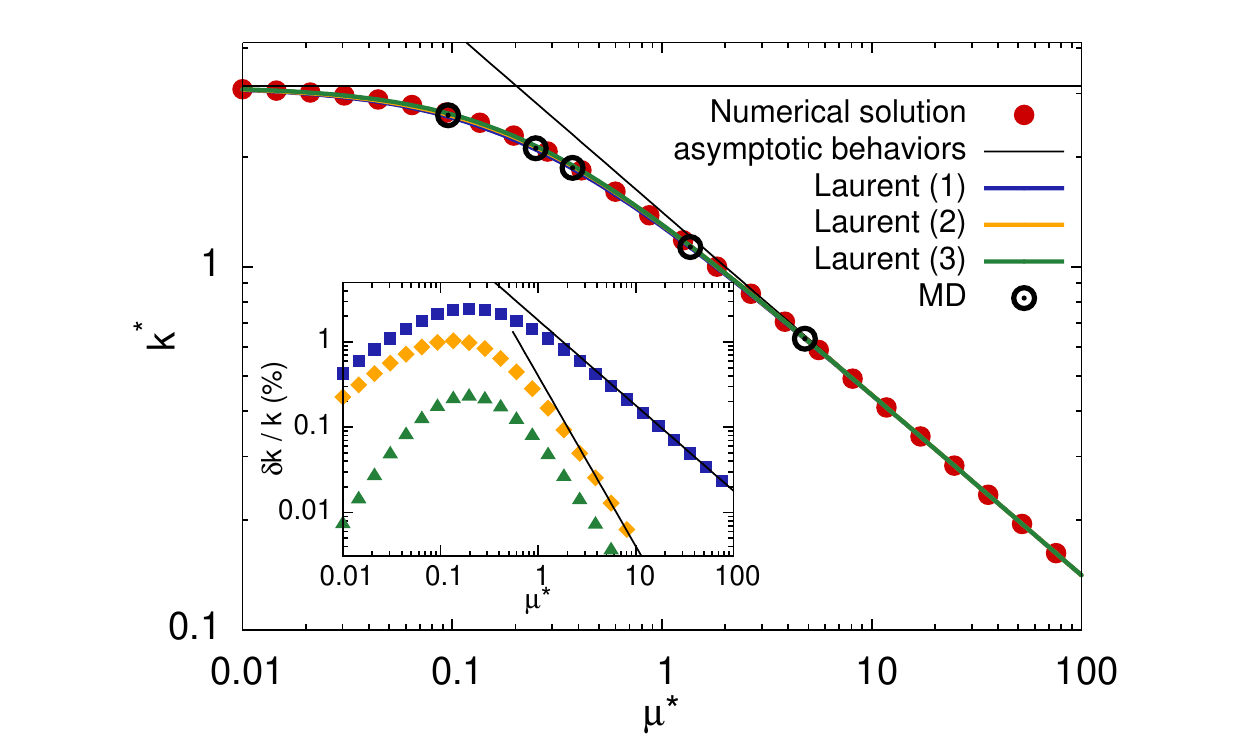}
\caption{Numerical solution of Eq.~(\ref{nonlin2})  for the the screening parameter $k^*$ (circles); MD results (open circles); Approximant Eq.~(\ref{ansatzK}) (thick solid lines); Asymptotic behaviors (thin lines). Inset: Relative error of the approximants (squares: Laurent order 1, Eq.~(\ref{ansatzK}); diamonds: Laurent order 2, Eqs.~(\ref{laurent}) and (\ref{Laurent2}); triangles: Laurent order 3, Eqs.~(\ref{laurent}) and (\ref{Laurent3}); thin lines: Asymptotic behaviors $1/\mu^{*}$ and $1/\mu^{*2}$). Note that $\delta k / k = \delta k^*/k^*$.}\label{fig1}
\end{figure}

We solved Eq.(\ref{nonlin2}) numerically, using a modified Powell's method\cite{Powell64} as provided by the MINPACK library\cite{more1984minpack}, for 100 different values of $\mu^*$, logarithmically distributed in the interval $[10^{-2}, 10^2]$. The resulting points $k^*(\mu^*)$ are reported in Fig.\ref{fig1}, along with the asymptotic behavior in the limits $\mu^*\to 0$ and $\mu^*\to \infty$, Eq.~(\ref{limit}).
 
To explicitly control the validity of these results, we performed molecular dynamics simulations of 240 ions distributed randomly on two walls and 240 counterions free to move between them. 

A Weeks--Chandler--Andersen  (WCA)
potential\cite{weeks71a} $$U(z)=4k_BT [(z/\sigma_{W})^{-12} - (z/\sigma_{W})^{-6}] + k_BT$$ 
for $z<2^{1/6}\sigma_{W}$, and $U=0$ otherwise, 
is acting between the confining walls and the 
mobile ions. The two walls are placed at a distance of $18\sigma_{W}$, so that ions can freely span a range $L\simeq 16 \sigma_W$ ($L^*=1$), and have a surface area  $S=24\sigma_{W}\times 24\sigma_{W}$ ($S^*=2.25$).
Mobile and fixed ions interact with each 
other only through the Coulomb interaction.
The electrostatic energy and the corresponding forces on ions are computed using the ELC algorithm\cite{dejoannis02a}, which takes into account the long range interaction between periodic copies only along the directions parallel to the confining surfaces.
In order to sample the canonical ensemble, we used a Langevin thermostat with integration timestep $\delta t = 0.001\sigma_{W}/\sqrt{k_BT}$, friction coefficient
$\gamma = 0.1\sqrt{k_BT}/\sigma_{W}$ and unit mass (note that the actual values of the timestep, the friction coefficient and the mass are irrelevant for the determination of equilibrium static quantities, as long as  the product $\gamma \delta t \ll 1$, and the microcanonical equations of motion are integrated accurately, which we tested by measuring an energy drift of about $3\times10^{-4}k_BT$/step). 

To make a real world example, one could consider silica walls separated by $L=20$~nm with surface charge $\sigma e= -0.32$~mC/m$^2$, corresponding to $\sigma\simeq 2\times 10^3 / \mu\mathrm{m}^2$ ($\sigma^*=0.8$), filled with water at T=300~K, which has a Bjerrum length $\lambda \simeq 0.7$~nm ($\lambda^*=0.035$). This would lead to a  Gouy-Chapman length $\mu\simeq 0.11$~$\mu\mathrm{m}$ ($\mu^*=5.5$). 

We have simulated the system with the ESPResSo simulation package\cite{limbach06a,arnold2013espresso} at different values of $\mu^*$, by changing the Bjerrum length $\lambda^*$, in the range $\mu^*$ = 0.09 -- 4.5.

\begin{figure}[t!]
\includegraphics[width=\columnwidth,trim={30 5 30 10},clip]{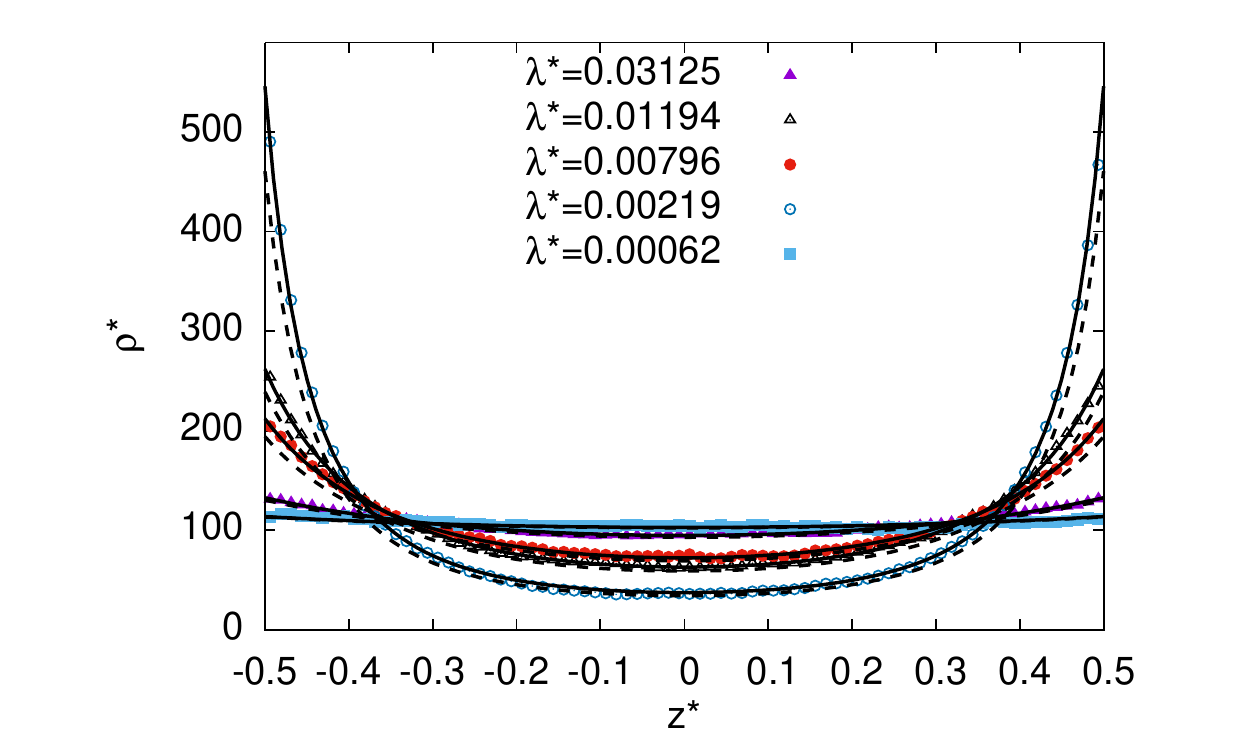}
\caption{Density profiles $\rho^*(z^*)$ for different values of $\lambda^*$, as obtained from MD simulations. 
Continuous lines are the result of the best fit to the Poisson--Boltzmann equation Eq.~(\ref{PBsol}). 
Dashed lines are the solutions obtained using the approximation Eq.~(\ref{ansatzR}).
The location of the walls is at $z^*=\pm 9/16\simeq\pm 0.563$, and ions can freely access the range $z^*\simeq[-0.5,0.5]$.}\label{fig2}
\end{figure}

\begin{table}[t!]
\caption{\label{table} Values of the screening parameter $k^*$ for different values of $\lambda^*$ (and corresponding $\mu^*$), as obtained from: the numerical solution of Eq.~\ref{nonlin2};  the approximant Eq.~(\ref{ansatzK});  MD simulation data by fitting Eq.~(\ref{PBsol}).}

\begin{tabular}{ccccc}
\multicolumn{1}{c}{$\lambda^*$} &\multicolumn{1}{c}{$\mu^*$} &  \multicolumn{1}{c}{$k^*$ (numer.)} & \multicolumn{1}{c}{$k^*$ (appr.)}& \multicolumn{1}{c}{$k^*$ (MD)}\\
\hline
0.03125 &    0.0955  & 2.646 & 2.590 & 2.612 \\
0.01194&    0.25    & 2.153 & 2.102 & 2.118 \\
0.00796&    0.3749  & 1.902 & 1.861 & 1.871 \\
0.00219&    1.3642  & 1.142 & 1.129 & 1.132 \\
0.00062&    4.7986  & 0.635 & 0.632 & 0.633 

\end{tabular}
\end{table}

The density profiles obtained with the MD simulations are shown in
Fig.~\ref{fig2}, together with the corresponding best fit curves to the
Poisson--Boltzmann solution, Eq.~(\ref{PBsol}) (solid lines), and the distributions obtained using the approximation~Eq.(\ref{ansatzR}) for the density in the middle of the channel.  The values of $k^*$
resulting from the best fit of MD data are reported in Fig.~\ref{fig1}
and Tab.~\ref{table}, showing very good qualitative agreement with
the numerical solution, despite the involved approximation of using a WCA potential for the walls.

In order to approximate the curve $k^*(\mu^*)$ , we used the following Ansatz: 
\begin{equation}
  k^* = \sqrt{\frac{2}{\mu^*+2/\pi^2}}\quad\mathrm{(approx.)},\label{ansatzK}
\end{equation}
or, equivalently, for the density in the middle of the channel,
\begin{equation}
\rho_0^* = \frac{2}{1/\sigma^*+4\lambda^*/\pi}=\frac{\sigma^*}{1/2+1/(\pi^2\mu^*)}\quad\mathrm{(approx.)}.\label{ansatzR}
\end{equation}
Eq.~(\ref{ansatzK}) satisfies both asymptotic behaviors, Eqs.~(\ref{limit}), and, automatically, also the constraint that at vanishing Bjerrum lengths the charge should be uniformly distributed. In the limit $\lambda^*\to 0$, in fact, the electrostatic interaction between ions is negligible with respect to thermal fluctuations at all scales and the system is expected to behave like an ideal gas, as in this case $\rho_0^*=NL^3/V = 2 \sigma L^2 = 2\sigma^*$, where $V=SL$ is the volume in which $N$ counterions are present.

Eq.~(\ref{ansatzK}) is plotted in Fig.~\ref{fig1}, where, on double logarithmic scale, it can barely be distinguished from the numerical data. We also report in the same figure the relative error $\delta k^*/k^*$ ($\delta k^*$ being the difference between the approximant and the numerical solution), which is always less than 2.5 and 5\% for  $k^*$ and $\rho^*$, respectively ($\delta \rho^*/\rho^* = 2 \delta k^* /k^* $). The error $\delta k^*/k^*$ displays a clear $1/\mu^*$ dependence,  showing that the next to the leading term in the asymptotic behavior of $k^*(\mu^*)$ is $\mathcal{O}(\mu^{*-1})$, as it is expected from Eqs.~(\ref{limit}). 

The value $\mu^*=2/\pi^2$ discriminates between two different regimes, which are characterized by the constant screening limit  $k^*\approx\pi$  (or, $\mu^*\ll 2/\pi^2)$, and by the dependence $k^*\approx \sqrt{2/\mu^*}$ (or, $\mu \gg 2/\pi^2$). For the reduced density $\rho^*$ with a fixed surface charge $\sigma^*$ the two regimes are, 
\begin{equation}
\begin{array}{rl}
\rho_0^*\approx 2 \sigma^* & \mathrm{if~} \mu\gg 2/\pi^2\\
\rho_0^*\approx  \pi^2  \sigma^* \mu^* & \mathrm{if~} \mu^* \ll 2/\pi^2.
\end{array}
\end{equation}

Our Ansatz does not bear any particular physical meaning, as far as we can 
say. However, it is interesting to notice that Eq.~(\ref{ansatzK}) (or better, its square) can be regarded as the first term of a Laurent series 
\begin{equation}
k^{*2}(\mu^*) = \sum_{n=1} \frac{b_n}{(\mu^*+\mu_0^*)^n}.
\label{laurent}
\end{equation}
The coefficients $b_n$ and $\mu^*_0$ can be determined by imposing the 
limiting behaviors, Eqs. (\ref{limit}), simultaneously, as we show in 
the Appendix, including the desired order. It is trivial to check that by considering only the first term in the expansion, and imposing that the  $\mu^{*0}$ and $1/\mu^*$ behaviors are satisfied, one recovers  Eq.~(\ref{ansatzK}).
If one takes instead into account further terms of the Laurent 
expansion, Eq.~(\ref{laurent}), it is possible to 
obtain more accurate approximations.  Two cases, in 
which the orders 
$\mu^{*0}$,$1/\mu^*$,
$1/\mu^{*2}$ 
and, additionally,
$\mu^*$,
are taken into account, are reported in 
Fig.~\ref{fig1}, and labeled  Laurent order 2 and Laurent order 3, respectively. 
The improvement is noticeable:
for the order 3 case, the relative error is always below approximately 0.2~\%, and 
decays to zero much faster, since the approximation is 
bound to match higher orders in the limiting behavior 
of the exact solution.
However, as one can see in the Appendix, the expressions for these 
approximations are more complicated and thus less
appealing than Eq.~(\ref{ansatzK}) for analytical manipulations.

The knowledge of the density at one of the plates allows one to compute the osmotic pressure $P$ thanks to the contact theorem\cite{henderson78,henderson79,mallarino15}, $ P^*/(2\pi\lambda^* \sigma^{*2}) = \rho^*(1/2) -1 $, where $P^*=L^3 \beta P$.  Noticing that at the contact (namely, $z^*=1/2$) one can write  $\tan(k^*/2)=\sqrt{1/\cos^2(k^*/2)-1}$ because $0<k^*<\pi$,  after some algebra one can reach the exact expression 
\begin{equation}
P^* = \mu^*\sigma^{*2}\left[ 1 + (k^*\mu^*)^2 \right]-\sigma^*/\mu^*,
\end{equation}
or, using Eq.~(\ref{ansatzK}), the approximation
\begin{equation}
P^*= \mu^*\sigma^{*2}\left( \frac{2\mu^{*2}}{\mu^*+2/\pi^2}+1\right)-\sigma^*/\mu^*\quad\mathrm{(approx.)}.
\end{equation}

\section{Conclusions}
We have presented a simple analytical approximation to the numerical solution of the nonlinear equation that determines the counterion density in the middle of a slit pore, in the Poisson--Boltzmann limit. 
This approximation for the salt-free, weak electrostatic coupling case is always within 5\% of the numerical solution and satisfies the asymptotic behavior at large and small Gouy--Chapman lengths of the screening constant and of the counterion density. 
This approximation can prove to be 
useful in estimating the full Poisson-Boltzmann 
solution of the distribution of ions 
between two charged plates with reasonable 
accuracy, without the need to solve the
associated nonlinear 
equation, and, for example, in performing 
analytical estimates of the complete solution dependence
on the parameters of the problem.  

\appendix
\section{Second and Third order Approximations}
Let us first expand the second asymptotic behavior in Eq.~(\ref{limit}) up to the second order. This can be done by recasting Eq.~(\ref{nonlin2}) in the form
\begin{equation}
k^*\mu^*=\cot(k^*/2)
\end{equation}
and expanding $k^*\approx k^*_0 + k^*_1 \mu + k^*_2\mu^{*2}$, where $\mu^*$ is the small expansion parameter. At order $\mu^{*0}$, we recover $\cot (k^*_0/2)=0$, or $k^*_0=\pi$. At order $\mu^{*}$, one obtains the equation $\pi \mu^*=\cot[(\pi+k_1^*\mu^*)]\approx-k_1^*\mu^*/2+\mathcal{O}(\mu^{*3})$, or, $k_1^*=-2\pi$. The same expansion for the cotangent gives access to the $\mu^{*2}$ term, $k_2^*=4\pi$. By taking the square of the expansion of $k^*$, one finally reaches the result
\begin{equation}
k^{*2} \approx \pi^2 - 4 \pi^2 \mu^* + 12\pi^2\mu^{*2} + \mathcal{O}(\mu^{*3}).\label{limit2}
\end{equation}
Let us now consider the Laurent expansion, Eq.~(\ref{laurent}), up to the term $n=3$, therefore setting all coefficients $b_4$, $b_5$,$\ldots$ to zero. 
In the limit $\mu^*\gg 1$, the expansion behaves as 
\begin{equation}
k^{*2} \approx \frac{b_1}{\mu^*} +\frac{b_2-b_1 \mu_0^*}{\mu^{*2}}+ \frac{\mu_0^* (b_1 \mu_0^*-2 b_2) +b_3}{\mu^{*3}}+\mathcal{O}(1/\mu^{*4})\label{expansion_inverse}
\end{equation}
The opposite limit, $\mu^*\ll 1$, instead, can be written as
\begin{align}
k^{*2}\approx\left( \frac{b_1}{\mu_0^*}+\frac{b_2}{\mu_0^{*2}} + \frac{b_3}{\mu_0^{*3}}\right)-\\
\mu^* \left(\frac{b_1}{\mu_0^{*2}}+2 \frac{b_2} {\mu_0^{*3}}+3 \frac{b_3}{\mu_0^{*4}}\right) +\mathcal{O}(\mu^{*2}) \label{expansion_direct}
\end{align}
The second order approximation presented in 
Fig.~\ref{fig1} can be obtained by setting $b_3=0$ and 
solving the system of equations for $b_1$, $b_2$ and 
$\mu^*_0$, with the requirement of a simultaneous 
matching of the coefficients of the asymptotic 
expansions for $k^{*2}$ at orders $\mu^{*0}$ and 
$1/\mu^*$. The coefficients are determined by the solutions of a second 
degree polynomial in $\mu^*_0$, and therefore two sets of solutions are present. Both behave asymptotically as imposed, but the result with the positive root,
\begin{equation}
\begin{cases}
b_1=2\\
\mu_0^* = 2/\pi^2+\sqrt{4/\pi^4-1/(3\pi^2)}\\
b_2=2\mu_0^*-1/3,\label{Laurent2}
\end{cases}
\end{equation}
yields the best approximation, and is reported in the inset of Fig.~\ref{fig1} with diamonds.

If the $b_3$ term is included, one has to match the 
coefficients of the asymptotic expansions including 
also the $\mu^{*}$ and $1/\mu^{*2}$ ones. The 
resulting system of equations has still an analytical 
solution (it is a third order polynomial in $\mu^*_0$), and the set of coefficients that yields the  best approximation is 
\begin{equation}
\begin{cases}
b_1=2\\
b_2=0.0632751\\
b_3=-0.0142314\\
\mu_0^*=0.198304.\label{Laurent3}
\end{cases}
\end{equation}
\\

\section*{Acknowledgments}
We thank Swetlana Jungblut for useful discussions. This work has been supported by the Marie-Sk\l{}odowska-Curie European Training Network COLLDENSE (H2020-MSCA-ITN-2014 Grant No. 642774).

\bibliographystyle{elsarticle-num}
\bibliography{biblio}

\end{document}